\documentclass[%
 reprint,
 amsmath,amssymb,
 aps,
]{revtex4-2}
\usepackage{braket}
\usepackage[caption=false]{subfig}
\usepackage{graphicx}
\usepackage{dcolumn}
\usepackage{physics}
\usepackage{amsmath}

\usepackage{soul,xcolor}
\setstcolor{blue}
\usepackage{bm}
\newcommand{\blue}[1]{\textcolor{black}{#1}}
\newcommand{\newblue}[1]{\textcolor{blue}{#1}}

\begin{document}

\title{Dynamics of vortex defect formation in two dimensional Coulomb crystals}

\author{Michael Arnold}\email{michael.arnold@sydney.edu.au}
 \affiliation{Centre for Complex Systems, Faculty of Engineering, The University of Sydney, Sydney, NSW 2006, Australia.}
\author{Ramil Nigmatullin} \email{ramil.nigmatullin@mq.edu}
 \affiliation{Center for Engineered Quantum Systems, Dept. of Physics \& Astronomy, Macquarie University, 2109 NSW, Australia.}

\date{\today}

\begin{abstract}
We study the non-equilibrium dynamics of two dimensional planar ion Coulomb crystals undergoing a structural buckling transition to a three plane configuration, driven by a reduction of the transverse confining frequency. This phase transition can be theoretically modeled using a mapping to a two dimensional Ginzburg-Landau theory with complex order parameter field. We demonstrate that finite rate quenches result in creation of stable topological vortices, which are localized point regions around which the phase of the order parameter field winds a multiple of $2\pi$. The density of the defects as a function of quench rate is investigated using molecular dynamics simulations and its scaling is shown to be consistent with Kibble-Zurek theory of defect formation. Following the quench, the annihilation of vortex and anti-vortex pairs results in the relaxation of defect density that follows a diffusive scaling with a logarithmic correction. This work highlights the potential for investigating complex non-equilibrium statistical physics of topological defects in an experimentally accessible ion trap setting. 
\end{abstract}

\maketitle

\section{Introduction}
Trapped ions is one of the most prominent quantum technologies. Doppler laser cooling can bring ions to low temperatures at which the ions crystallize forming regular structures, known as Coulomb crystals, whose shape is determined by the trapping parameters. 
Linear chain crystals in Paul traps have the simplest phonon spectrum and have been widely used in metrology \cite{PhysRevApplied.11.011002}, quantum computing \cite{HAFFNER2008155,doi:10.1063/1.5088164} and quantum simulations \cite{Kim2010}. In Penning traps, large two dimensional planar crystals can be readily created and these structures have also been used in metrology and quantum information processing applications \cite{Britton2012}. 

Beyond these simple Coulomb crystal geometries, there is great interest in exploring more complex structural phases and the transitions between them \cite{RevModPhys.71.87,PhysRevB.81.024108,PhysRevB.77.064111,PhysRevB.90.094111,Cornelissens2000}.
 The strong long-range Coulomb interactions between particles leads to highly non-linear non-trivial dynamics, whose investigation is of fundamental interest in the fields of nonlinear science, complex systems and solid-state physics; it is useful as a platform for studying complex non-linear and non-equilibrium dynamics in areas including the simulation of Klein-Gordon fields on a lattice \cite{PhysRevLett.101.260504}, Kibble-Zurek mechanism of defect formation \cite{Pyka2013,Ulm2013,PhysRevB.93.014106}, dynamics of discrete solitons \cite{Partner_2013,Landa_2013,PhysRevLett.110.133004}, dry friction \cite{Kiethe2017}, energy transport \cite{Pruttivarasin_2011,PhysRevResearch.2.033198} and synchronization \cite{PhysRevLett.106.143001}. Coulomb crystals with more complexity also provide new lattice geometries for quantum simulations \cite{PhysRevLett.107.207209}. 


In this paper, we investigate numerically the non-equilibrium structural phase transition from a quasi-two-dimensional 1-plane crystal to a 3-plane crystal. We focus on the Kibble-Zurek mechanism of the formation of topological defects and the subsequent coarsening dynamics of annihilation of defects and anti-defect pairs. 

Previous studies of KZ mechanism in ion crystal systems have focused on the linear to zigzag phase transition in a quasi-one-dimensional system \cite{Pyka2013,Ulm2013}. This are symmetry breaking phase transitions and the resulting defects are either kinks \cite{Pyka2013,Ulm2013} if the $Z_2$ symmetry is broken or helical twists \cite{PhysRevB.93.014106} if the $U(1)$ symmetry is broken. \newblue{The non-equilibrium $U(1)$ symmetry breaking leading to stable winding of the order parameter has also been studied in one dimensional BECs in toroidal traps \cite{Das2012}.} 

The 1-plane to 3-plane structural phase transition considered in this paper can be described by an XY 6-clock model with an intermediate Kosterlitz-Thouless phase. The defects are $U(1)$ point vortices whose physics is considerably richer \cite{PhysRevX.6.031025}.
Previously, simulations of finite quenches in a two dimensional XY spin model have shown that the density of vortices is dictated by both the KZ mechanism and coarsening dynamics of annihilation of vortex/anti-vortex pairs \cite{Jelic2011}. This observation was corroborated in an experimental study of colloids undergoing a phase transition via Kosterlitz– Thouless–Halperin–Nelson–Young (KTHNY) melting scenario \cite{Deutschlaender2015}. 
\blue{Our molecular dynamics simulations demonstrate that faster quenches result in higher densities of created defects in qualitative agreement with KZ theory. At late times of the quench protocol the defects annihilate through coarsening dynamics. The density of defects eventually stabilizes, which may be attributes to the pinning effect of the emergent 6-clock potential.}
Our work introduces a new platform for investigating the non-equilibrium KT phase transition and more generally the collective dynamics of interacting vortices.

The paper is organized as follows. Section \ref{sec:GLmodel} presents the microscopic model, reviews its mapping to the Ginzburg-Landau field theory and the phase diagram of a 1-plane to 3-plane structural transition. Section \ref{sec:defects} uses molecular dynamics simulations to demonstrate the existence of topological defects in the 3-plane phase and presents an algorithm for determining their location. Section \ref{sec:KZscaling} focuses on molecular dynamics investigation of finite rate quenches, where the KZ mechanism and coarsening dictate the evolution of the average number of defects in the system. 



\section{Ginzburg-Landau model of 1- to 3-plane transition} \label{sec:GLmodel}

Ion traps confine repulsively interacting ions in space either by rapidly varying oscillatory electric fields, as in the Paul traps, or a combination of static electric and magnetic fields, as in the Penning traps \cite{Gosh}. 
The dynamics of ion Coulomb crystal can be often approximated with the so-called pondermotive approximation or pseudopotential theory, which replaces the time-varying trapping fields experienced by particles by an effective time-independent harmonic potential. 
The laser cooling reduces the temperature of the ions such that they can form regular crystal-like configurations, whose overall shape is determined by the trap parameters.
We will consider a system of $N$ ions confined to a periodic cell in the $x$-$y$ plane and by the harmonic confinement in the $z$-direction. The potential energy is given by

\begin{equation}
    V = \frac{1}{2} m\omega_z \sum_j^N z_j^2 + \mathcal{K} \sum_{i<j}\frac{1}{|\textbf{r}_i-\textbf{r}_j|} \label{eq:pe}
\end{equation}
where $\textbf{r}_j = (x_j,y_j,z_j)$ are the coordinates of the $j$th ion, $\mathcal{K}\equiv q^2/4\pi\epsilon_0$, $q$ is the charge of the ion, $\epsilon_0$ is the vacuum permittivity, $m$ is the mass of the ion and $\omega_z$ is the trapping frequency in the $z$-direction. The periodic boundary conditions results in a homogeneous spacing in the ion crystal. In a real experimental system, the open boundary conditions and the harmonic confinement in the $x$ and $y$ direction would result in inhomogeneous spacing between the ions; the ions are closer together in the centre of the crystal and further apart near the edges. The system with periodic boundary condition can be viewed as an approximation to a central region of a large ion crystal, where the spacing is approximately homogeneous and the boundary effects can be neglected. 

\begin{figure}
    \centering
    \includegraphics[scale=0.6]{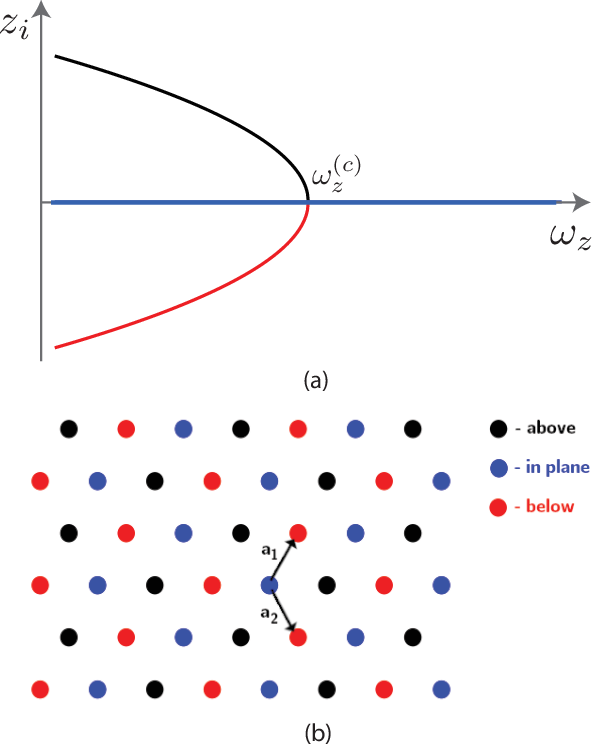}
       \caption{Structural buckling transition between a 1-plane and 3-plane configuration. (a) The transverse displacement of the ions in a crystal at different values of $\omega_z$ in the vicinity of the critical $\omega_z^{(c)}$. (b) The triangular lattice structure of the 3-plane phase.}
         \label{fig:bifurcation}
\end{figure}

Above a certain critical value of $\omega_z=\omega_z^{(c)}$ the lowest energy configuration is a planar triangular lattice. 
When the confining frequency is reduced to below $\omega_z^{(c)}$, the 1-planar crystal configuration undergoes a buckling structural transition into a 3-planes, all of which in triangular lattice geometry but with double the lattice spacing (see Figure \ref{fig:bifurcation}). This buckling instability has been predicted in an early theoretical work by Dubin \cite{PhysRevLett.71.2753} and has been observed experimentally in \cite{Mitchell1998}. Recently, Podolsky \emph{et. al.} \cite{PhysRevX.6.031025} derived a Ginzburg-Landau (GL) field theory for this transition thereby proving that it is in the universality class of a two dimensional XY model \cite{PhysRevX.6.031025}.

The GL field theory is derived by Taylor expanding the non-linear Coulomb interaction term in equation (\ref{eq:pe}) in displacements around the equilibrium lattice positions. In \cite{PhysRevX.6.031025} it was shown that one must keep the terms up to the sixth order in the expansion to correctly capture the critical properties of the structural phase transition. The GL free energy density is give by

\begin{equation}
    \frac{f}{\mathcal{K}}= \frac{\gamma}{2} |\nabla\psi|^2+ \epsilon |\psi|^2+u|\psi|^4+v|\psi|^6+\frac{w}{2}\left[\psi^6 + (\psi^*)^6\right] \label{eq:GL}
\end{equation}
where $\epsilon=\frac{1}{\sqrt{3}}\left(\frac{m\omega_z^2 a^2}{2\mathcal{K}}-I_2\right)$, $u=3/\sqrt{3}/4 I_4$, $w=\frac{5}{8\sqrt{3}}I_6$, $v=-\frac{25}{4\sqrt{3}}I_6$, $I_2=6.683$, $I_4=3.56$, $\gamma=0.223$, $I_6=2.558$ and $a$ is the lattice spacing. The order parameter field at a lattice point with coordinates $(x_j,y_j$, $\psi(x_j,y_j)$, is an implicit function of the transverse displacement

\begin{equation}
    z_j = \textrm{Re}\left[\psi e^{i\textbf{K}\cdot\textbf{r}_j}\right]. \label{eq:phiDef}
\end{equation}
Here $\textbf{K}$ is the base vectors of the first Brillouin zone of the triangular lattice given by $\textbf{K} = (4\pi/3,0)$, $\textbf{r}_i = n_1 \textbf{a}_1 + n_2\textbf{a}_2$, with $\textbf{r}_{1,2} = (\frac{1}{2},\pm\sqrt{3}/2)$. The order parameters $\psi$ is complex and can be expressed as $\psi=|\psi|e^{i\theta}$. 


\begin{figure*}
    \centering
    \includegraphics[scale=0.6]{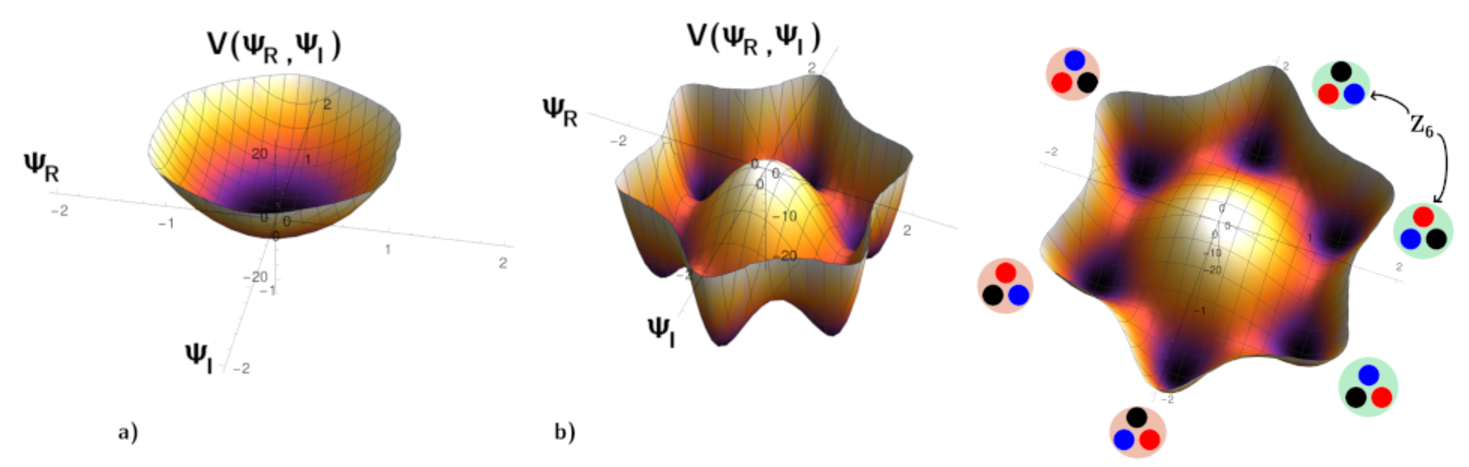}
     \caption{(a) Single well potential of GL theory for $\epsilon>0$ (b) Mexican hat potential for $\epsilon<0$.}
     \label{fig:mexicanhat}
\end{figure*}

\begin{figure}
    \centering
    \includegraphics[scale=0.8]{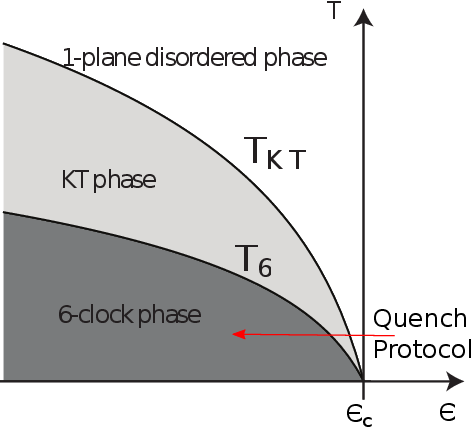}
    \caption{Phase diagram of a structural 1-plane to 3-plane phase transition of a Coulomb crystals. The control parameter space determined by the temperature $T$ and coefficient $\epsilon$ which is a function of the ratio $m\omega_z^2a^2/\mathcal{K}$. The line $T_{KT}$ indicates a transition between 1-plane phase and a 3-plane phase in quasi-long-range ordered KT phase. The line $T_6$ indicates a transition between the KT phase and the long range ordered 3-plane phase with the $Z_6$ symmetry.  \blue{The quench is implemented by reducing the transverse confining frequency $\omega_z$ at a rate $(\omega_z^{(i)}-\omega_z^{(f)})/\tau_Q$}}
    \label{fig:phasedia}
\end{figure}

The potential energy for the mean field configuration, which neglects the spatial fluctuations in the order parameter, $V(\psi)/\mathcal{K}=\epsilon|\psi|^2+u|\psi|^4+v|\psi|^6+\frac{w}{2}\left[\psi^6+(\psi^*)^6\right]$, is shown in Figure \ref{fig:mexicanhat}. 
For $\epsilon>0$ the order parameter is zero, $\psi=0$, and the system is in the 1-plane phase. The potential is a single well and since the order parameter has no preferred direction, the 1-plane phase is disordered. 
For $\epsilon<0$ the order parameter is non-zero and the system is in the 3-plane phase.  The potential is a Mexican hat but with 6 equally spaced wells which correspond to the local order of the 6 degenerate lattice arrangements shown in Figure \ref{fig:mexicanhat}(b). This corresponds to the 6-clock phase which has the discrete $Z_2\times Z_3$ symmetry. At higher energies the order parameter can easily overcome the energy barrier between the neighboring minima, and the symmetry changes to the broken $U(1)$ continuous symmetry. Two dimensional systems with the broken $U(1)$ symmetry in the order parameters support topological defect vortex configuration, which are localized regions where the field winds around the Mexican hat potential. The presence of these topological defects drastically alters the physics of the system leading to the existence of the KT phase  \cite{Berezinsky:1970fr,Kosterlitz_1973}. Thus, near the critical point of the 1-plane to 3-plane structural phase transition the system can exist in 3 phases, disordered, KT and the 6-face clock phases, depending on the value of $\epsilon$ and temperature $T$. The phase diagram was derived in \cite{PhysRevX.6.031025} and is sketched in Figure \ref{fig:phasedia}. The KT phase is characterized by a change in behavior in correlation length. For $T>T_{KT}$, the system is disordered, the correlation length decays exponentially and there is a finite density of unbound vortices. For $T<T_{KT}$, there is a quasi-long range order with a power law decays of the correlation length and vortices and anti-vortices form bound pairs. For $T<T_6$, the system is in the 6-clock phase, which again exhibits a long-range order with exponentially decaying correlation length.

\section{Topological defects in the 3-plane phase} \label{sec:defects}
To verify the prediction of the existence of the topological defects, we have performed molecular dynamics simulations of ion Coulomb crystals confined in a box with periodic boundary conditions in the $x$ and $y$ direction. The topological defects are produced by quenching a system from a 1-plane disordered phase into the 3-plane phase. We use a Langevin thermostat to simulate the interaction of the ions with the cooling laser beam, which thermalized the ions. The equations of motion for the $j$th ion are given by

\begin{eqnarray}
 m \partial_{tt} x_j & = & - m\gamma \partial_t x_j -\partial_{x_j} V_c + \theta_{xj}(t) \label{eq:motion1}\\
 m \partial_{tt} y_j & = & - m\gamma \partial_t y_j -\partial_{y_j} V_c + \theta_{yj}(t) \label{eq:motion2}\\
 m \partial_{tt} z_j & = & -m\omega(t)^2- m\gamma \partial_t z_j -\partial_{z_j} V_c + \theta_{zj}(t) \label{eq:motion3},
\end{eqnarray}
where $m$ is the mass of the ion, $\omega(t)$ is the transverse confining frequency, $V_c$ is the Coulomb interaction energy, $\gamma$ is the damping coefficient. The force $(\theta_{xj}, \theta_{yj},\theta_{zj})$ is is the stochastic thermal force satisfying $\langle \theta_{\alpha,j}(t) \rangle = 0$ and $\langle \theta_{\alpha,j}(t) \theta_{\beta,k}(t') \rangle = 2 m\gamma k_B T \delta_{\alpha\beta} \delta_{jk} \delta(t-t')$, where $\langle ...\rangle$ denotes ensemble averaging. \blue{The transverse frequency is varied linearly with time $\omega(t)=(\omega_z^{(i)}-\omega_z^{(f)})/\tau_Q t+\omega_z(i)$, where $\omega_z^(i)$ and $\omega_z^(f)$ are the initial and final frequencies.}  The integration of the equation of motion is performed using GPU accelerated OpenMM \cite{10.1371/journal.pcbi.1005659} framework, and Ewald sums are used to approximate the Coulomb interactions in the $x$ and $y$ directions. 

To determine the location of defects in a given ion crystal configuration, one must compute the local order parameter field $\psi$ using the individual ion coordinates. Using equation (\ref{eq:phiDef}) the order parameter in the 6-clock phase can be written as 
\begin{eqnarray}
z_i & = & \textrm{Re}[\Psi e^{\textbf{K}\cdot \textbf{r}_i}] \\
& = & |\psi|\cos\left(\frac{\pi(2n_i+1)}{6}+\delta \Theta_i +\textbf{K}\cdot\textbf{r}_i\right),\label{eq:orderpar2}
\end{eqnarray}
where $n_i\in \{1,...,6\}$ determines the clock state at the position of the $i$th ion and $\delta\Theta_i$ is the fluctuation about this phase. Denoting the splitting between the planes as $h\equiv\textrm{max}_i(|\langle z_i \rangle|))$, where $z_i$ is the $z$-coordinate of an ion either in the + or - sublattice of the 3-plane structural phase, one finds that $h=|\psi|\cos(\pi/6)\langle \cos(\delta\Theta)\rangle$ and equation (\ref{eq:orderpar2}) can be written as 

\begin{equation}
z_i = \frac{h}{\cos(\pi/6)} \cos\left(\delta\Theta_i+\frac{\pi(2n_i+1)}{6}+ \textbf{K}\cdot\textbf{r}_i\right). \label{eq:orderparam3}
\end{equation}

The values of clock state at each point, $n_i$, are determined by allocating a phase value to an arbitrary chosen patch of three adjacent ions and then assigning all other patches the best matching value relative to this chosen reference. After assigning $n_i$, the correction term, $\delta\Theta_i$, is obtained by solving numerically the non-linear equation (\ref{eq:orderparam3}) using gradient descent algorithm. 


\begin{figure*}
    \centering
    \includegraphics[scale=0.85]{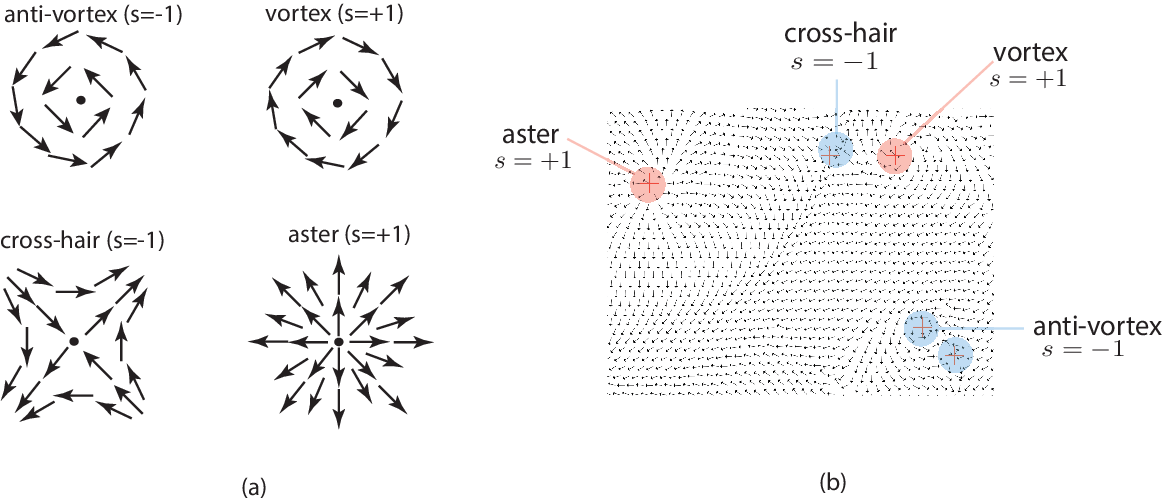}
    \caption{\blue{a) Topological defects in 2d XY model. Vortices and asters have topological charge $s=+1$. Anti-vortices and cross-hairs have topological charge $s=-1$ \cite{Shankar2021}. The field lines of asters are oriented radially. Vortices and anti-vortices have the order parameter field lines that are oriented orthoradially. The order parameter field of cross hairs does not have a $U(1)$ symmetry but exhibits a $\pi$ rotational symmetry. (b) Example of a order parameter field computed from an ion Coulomb crystal configuration after a finite rate quench from a single plane to a three plane structural phase.}}
    \label{fig:vortex_configuration}
\end{figure*}

Figure \ref{fig:vortex_configuration} shows a typical configuration with topological defects.  The locations of the defects are determined by finding localized regions on the boundary of which the phase $\Theta$ winds an integer multiple of $2\pi$. The topological charge $s$ of a defect is defined as a winding number along the contour $C$ encircling the defect i.e. $s = \frac{1}{2\pi} \int_C \partial_l \Theta dl$ where $l$ is the position along the path of the chosen contour. In our simulations, we observe 4 types of point defects: asters and vortices with charge +1, and cross-hairs and anti-vortices with charge -1. 
The existence of such defects may be predicted on general homotopy theoretic arguments \cite{RevModPhys.51.591}.
Here, we have demonstrated that in 3-plane Coulomb crystal all four types of defects are energetically stable.

\begin{figure*}
    \centering
    \includegraphics[scale=0.6]{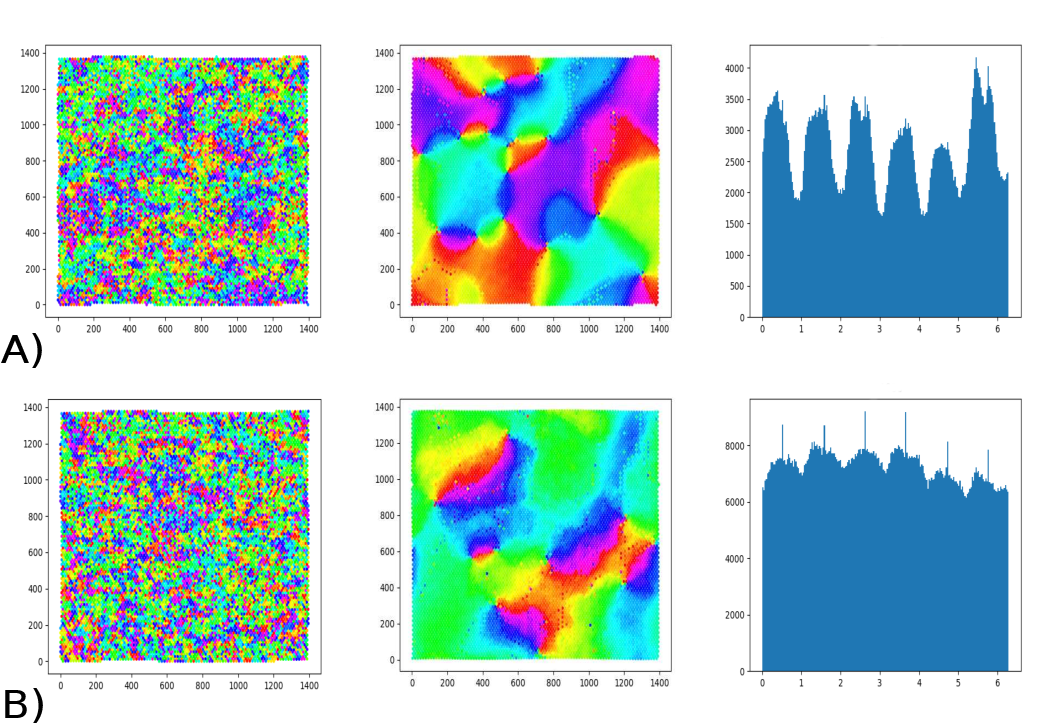}
    \caption{Instantaneous configurations of a quasi-two-dimensional ion crystal with 9858 $\textrm{Be}^{+}$ ions at a temperature of $20 \mu$K and a) $\gamma = 8.96 \times 10^{-7} \textrm{ps}^{-1}$ and b) $\gamma = 8.96 \times 10^{-8} \textrm{ps}^{-1}$ in the 3-plane structural phase at the end of a non-equilibrium quench from a disordered single plane phase at different rate $\tau_Q^{-1}$. The quench is implemented by reducing the transverse confining frequency $\omega_z$. The density of topological defects increases with increasing quench rate. Each color corresponds to a phase angle $\Theta$ of the local order parameter $\psi=|\psi| e^{i\Theta}$.  The first column demonstrates the disordered phase.  The second column shows vortexes in the clock phase.  The last column shows a histogram demonstrating a preference for of the six angles in the clock model. \blue{The histograms demonstrate less preference for the the six angles in the clock model with a lower gamma which is consistent with being in the KT phase.}}
    \label{fig:vortex_configuration_histogram}
\end{figure*}

\begin{figure*}
    \centering
    \includegraphics[scale=0.6]{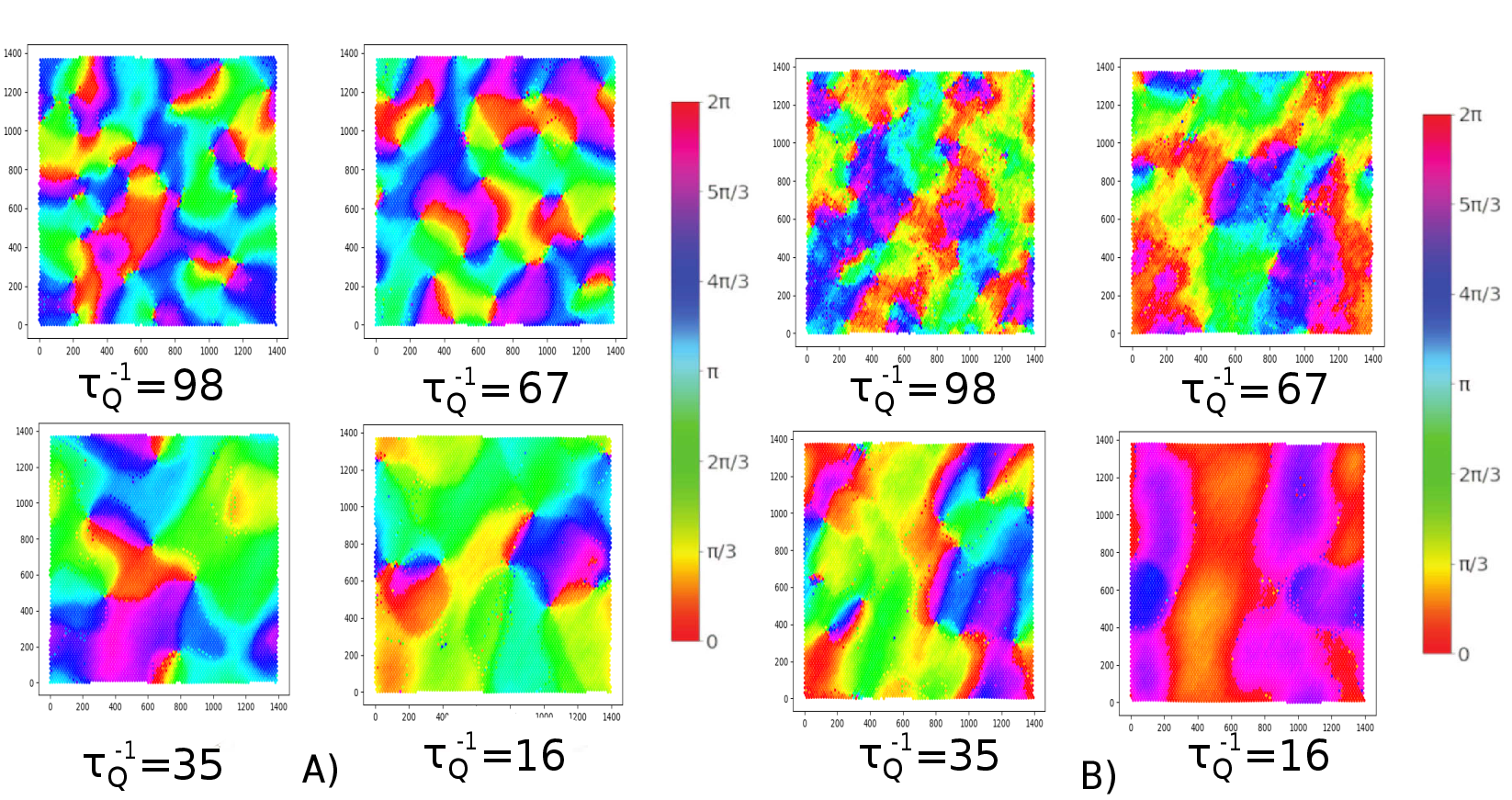}
    \caption{Examples of configurations with topological defects in the 3-plane structural phase represented as color plot. The observed defects are asters and vortices (topological charge, $s=+1$) or cross-hairs and anti-vortices (topological charge, $s=-1$) a) demonstrates configurations with $20 \mu$K and a) $\gamma = 8.96 \times 10^{-7} \textrm{ps}^{-1}$ and b) demonstrates configurations with $\gamma = 8.96 \times 10^{-8} \textrm{ps}^{-1}$}
    \label{fig:vortex_configuration_Tq}
\end{figure*}

\blue{The quench from a 1-plane to 3-plane structural configuration traverses two transitions, disordered to KT phase and KT phase to 6-clock phase, as shown in figure \ref{fig:phasedia}. The 6-clock phase exhibits long range order where the phase of the order parameter $\theta$ has preference toward the six angles that correspond to the ground state configurations. Thus, in order to establish whether the system is in KT or 6-clock phase, we evaluate the histograms of the angles $\theta(t)$ obtained from multiple simulation runs. The MD simulations are carried out at fixed temperature $T$, fixed starting and ending transverse confining frequencies, but at two different values of damping coefficient $\gamma$ and several values of quench times $\tau_Q$.  At high damping $\gamma$ the histogram of $\theta$ at the end of the quench displays six prominent peaks as shown in figure \ref{fig:vortex_configuration_histogram}A). This indicates that the system ends deep in the ordered 6-clock phase. At lower $\gamma$ the peaks are still present but their height is much smaller and the distribution of $\theta$ is close to uniform. This suggest that in this regime the system is closer to the quasi-long range ordered KT phase. Since the lower damping rate results in slower rate of dissipation of kinetic energy, the effective temperature during the non-equilibrium evolution is higher and thus the system samples the KT phase for a larger period of time.}

\blue{Figure \ref{fig:vortex_configuration_Tq} shows the vortex configurations at the end of the quenches at different rates $\tau_Q^{-1}$ with low (a) and high (b) damping coefficient. The higher quench rates result in a higher defect density, which is qualitatively consistent with the KZ mechanism of defect formation. The low quench rates result in a significantly more disordered order parameter field indicating that in this dynamical regime the system is closer to the quasi-long range ordered KT phase. In the rest of the paper, we focus on quantitatively probing the evolution of the average number of vortices and its dependence on the quench rate.}

\section{Kibble-Zurek mechanism of defect formation} \label{sec:KZscaling}
 The relationship between the number of defects and the quench rates across symmetry breaking phase transitions was initially investigated by Kibble in the context of cosmology \cite{Kibble1980} and Zurek in the context of condensed matter \cite{Zurek1985}, in what became known as Kibble-Zurek (KZ) theory. Experimentally, the KZ mechanism for a 6-clock model has been investigated in the context of ferroelectric materials \cite{PhysRevX.2.041022,Lin2014}. One should note, however, that in ferroelectrics the transition happens in three dimensions where there is no KT phenomena. 




Lets review the KZ mechanism as applied to continuous second order phase transitions. Consider approaching the critical point of a symmetry breaking second order phase transition. The correlation length $\xi$, defined by $\langle \psi(0,t)\psi(r,t)\rangle \sim e^{-r/\xi}$, diverges as a power law of the control parameter $\xi=\xi_0/|\epsilon|^\mu$, where $\mu$ is the critical exponent. The system also slows down on the approach to the critical point i.e. the relaxation time, $\tau$, defined as $\langle \psi(0)\psi(t)\rangle\sim e^{-t/\tau}$ diverges as $\tau = \tau_0/|\epsilon|^{\mu}$. KZ mechanism proposes that the correlation length freezes out when the relaxation time is equal to the time left until the crossing of the critical point i.e. the freeze out time $\hat{t}$ is found by solving $\tau(\hat{t})= \tau_Q$. This ``cross-over" time $\hat{t}$ marks a transition between the adiabatic dynamical regime, where the correlation lengths adjust to its equilibrium value, and impulsive regime, where the correlation length is fixed. For a linear quench $\epsilon = t/\tau_Q$, one observes $\hat{t} = (\tau_0 \tau^{\mu}_Q)^{1/(1+\mu)}$ and the freeze-out correlation length is $\hat{\xi} = \xi(\hat{t}) = \xi_0 (\tau_q/\tau_0)^{\nu/(1+\mu)}$. In the two dimensional system, the number of defects $n$ is inversely proportional to the square of the correlation length scale in the system i.e. $n\propto \xi^{-2}$. Thus, the KZ prediction for the defect density in the end of the quench is $n_f = \hat{\xi}^{-2}\sim (\tau_q/\tau_0)^{-2\nu/(1+\mu)}$.

For KT phase transition, the same arguments applies except the correlation length diverge exponentially at the critical point rather than algebraically i.e. 

\blue{
\begin{equation}
\xi = A e^{a |\epsilon|^{-\mu}} \label{eq:xiKT}
\end{equation}
\begin{equation}
\tau = B e^{b |\epsilon|^{-\nu}} \label{eq:tauKT}
\end{equation}
}
\blue{The freeze-out time, $\hat{t}$, is obtained by solving the equation, $\tau(\hat{t})=\tau_Q$}
\blue{
\begin{equation}
    e^{b|\hat{t}/\tau_Q|^{-\nu} = \tau_Q},
\end{equation}
which does not have a closed form solution. The freeze-out correlation length is given by
}
\blue{
\begin{equation}
    \hat{\xi} = \textrm{exp}(a |\epsilon(\hat{t})|^{-\mu}),
\end{equation}
Thus for the non-equilibrium transition between disordered and KT phase transition, there is no simple power-law dependence of $\hat{\xi}$ on the quench rate $\tau_Q^{-1}$. The number of defects, $n_d$, will have a complex functional dependence on $\tau_Q$, the universal scaling exponent $\mu$ and $\nu$ as well as the non-universal parameters $a$, $b$, $A$ and $B$. The quantitative verification of the KZ theory relies on first measuring the equilibrium scaling relations given by equations (\ref{eq:xiKT}) and (\ref{eq:tauKT}), as was previously done in the experimental study of KT scaling in colloids \cite{Deutschlaender2015}. 
}

\blue{The high mobility of vortices in two dimensional systems results in significant annihilations between vortex and anti-vortex pairs, adding to the complexity of studying the KZ mechanism in the KT systems. }
 Thus, the number of defects in the ``impulsive" regime is not fixed but continuously decreases as the defects annihilate. In the context of sudden quenches, this \blue{growth} of correlation length is known as coarsening. 

\newblue{The dynamics of defects modifies the observed KZ scaling laws for in both classical and quantum systems \cite{PhysRevB.104.014406}}.
In \cite{PhysRevE.81.050101,Jelic2011}, the authors propose that for a quench into the KT phase, one should account for both KZ mechanism and coarsening expressing the time evolution of correlation length as



\begin{figure}
\centering
    \includegraphics[scale=0.5]{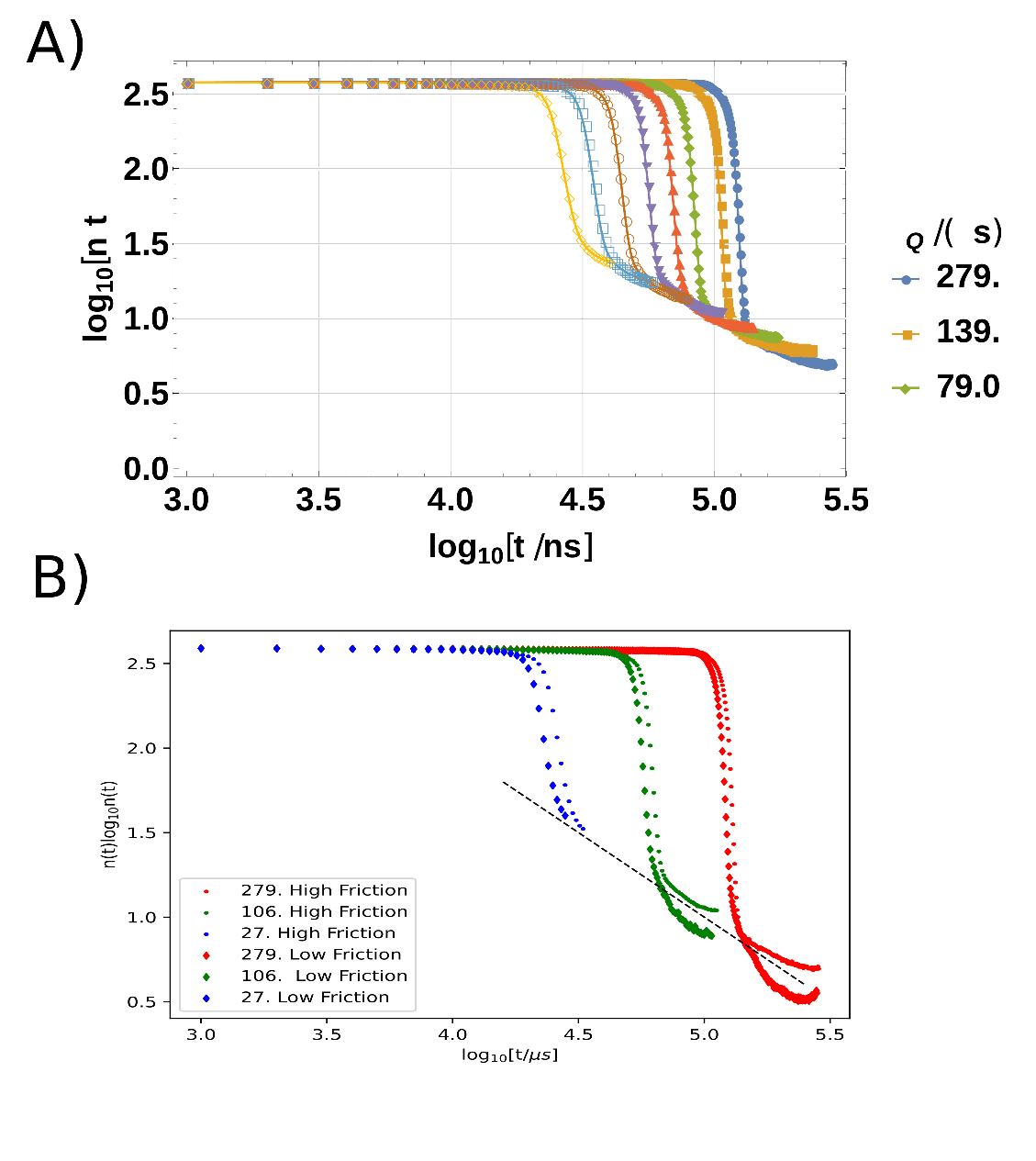}
  \caption{Evolution of the average number of defects following quenches from 1-plane disordered phase into a 3-lane 6 face clock phase at finite rate $1/\tau_Q$. The simulated system contained 9858 $\textrm{Be}^{+}$ ions at a temperature of $20\mu$K while \blue{$\gamma = 8.96 \times 10^{-7}\textrm{ps}^{-1}$ in the high friction case, and $\gamma = 8.96 \times 10^{-8}\textrm{ps}^{-1}$ in the low friction case. The ions are} confined in a periodic box in the $xy$ directions of size $1395.00\times1376.98$ $\mu$m. The starting and ending frequencies were set to $\omega_z^{(i)}=7.70\textrm{ Mhz}$ and $\omega_z^{(f)}=7.42\textrm{ MHz}$, and the critical frequency is $\omega_z^{(i)}=7.60\textrm{ MHz}$.  \blue{A) is the evolution of the number of defects in the high friction case, while B) shows both the high friction and low friction cases on the same plot.}}
  \label{fig:nvst} 
\end{figure}

\begin{figure}
\centering
    \includegraphics[scale=0.5]{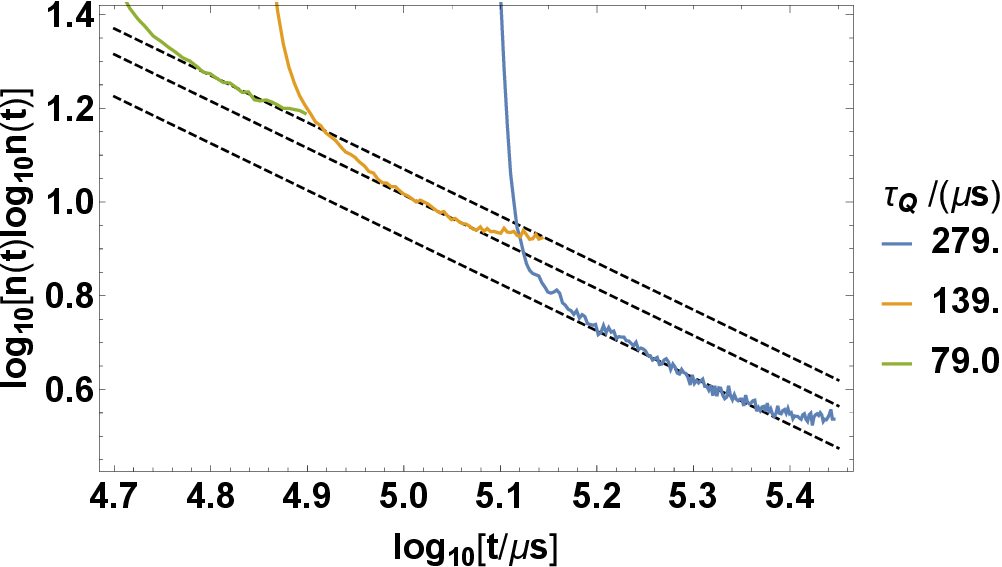}
  \caption{Late time evolution of the average number of defects for three quenches at different rates $1/\tau_Q$. The slopes of the dashed lines is -1, indicating a $n\ln n \sim t^{-1}$ coarsening scaling. }
  \label{fig:coarsening} 
\end{figure}

\begin{figure}
\centering
    \includegraphics[scale=0.6]{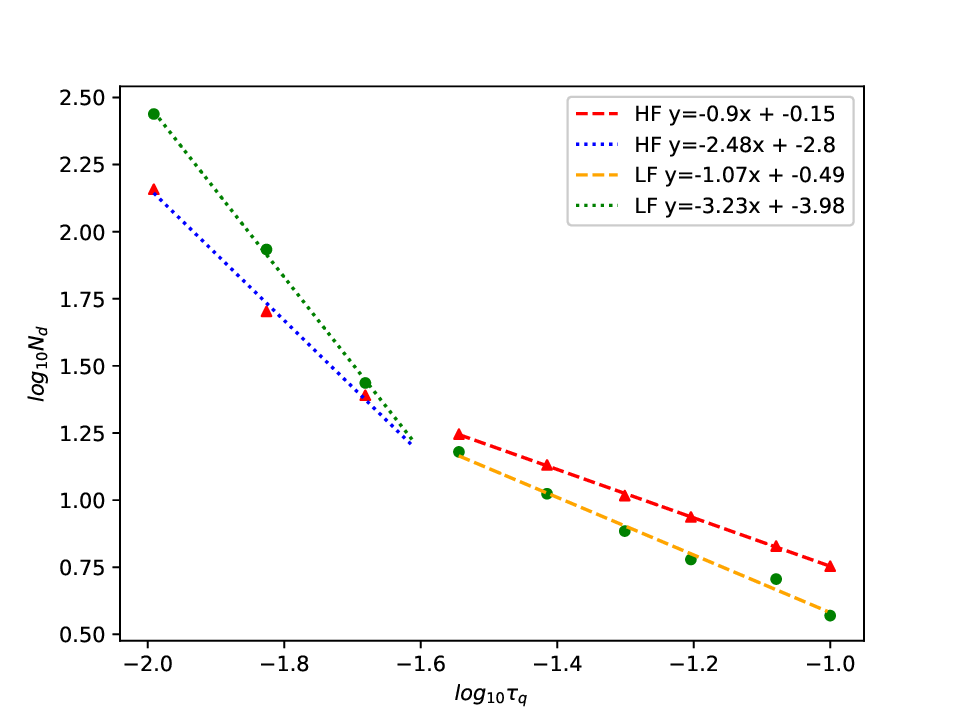}
  \caption{Plot of the number of defects at the end of the quench, $n_f$ as a function of quench rates with high friction (HF) $\gamma = 8.96 \times 10^{-7}\textrm{ps}^{-1}$ and low friction (LF) $\gamma = 8.96 \times 10^{-8}\textrm{ps}^{-1}$.}
  \label{fig:lastN}
\end{figure}


\begin{align}
\xi(t) 
     &= \begin{cases}
         \xi_{eq}(t),
         & \textrm{for }t <\hat{t}
         \\
        \hat{\xi} + f(t),
        & \textrm{for }t \geq \hat{t}        \end{cases}, \label{eq:KZcoarsening}
\end{align}
where $f(t)$ is the function representing the growth of the correlation length due to coarsening. Equation (\ref{eq:KZcoarsening}) expresses the idea that before the KZ freeze-out time, $\hat{t}$, the correlation length adopts its thermal equilibrium value, and after crossing $\hat{t}$ the correlation length is growing via coarsening. In a two dimensional system, which is quenched from disordered into ordered phase in the presence of linear damping, one expects the coarsening to proceed via diffusing law, where the correlation length grows as the square root of time $f(t) \sim t^{1/2}$ \cite{PhysRevA.42.5865}. Several studies noted that the approach to diffusive law can be slow and that the coarsening is more accurately described by including a logarithmic correction, $f(t) \sim \left(t/\ln t\right)^{1/2}$\cite{PhysRevE.47.1525,PhysRevLett.84.1503}. 

\blue{We have carried out an MD simulation to probe the KZ and coarsening dynamics following a dynamic crossing of the 1-plane to 3-plane structural phase transition}
 We simulate the dynamics of $N=9858$ $\textrm{Be}^{+}$ ions at a temperature of $20\mu \textrm{K}$  confined to a periodic box in an $xy$ directions and a harmonic potential in the $z$ directions by numerically integrating equation (\ref{eq:motion1})-(\ref{eq:motion3}). \blue{To establish whether the collective dynamics is sensitive to the friction coefficient, the simulations are carried out at high damping, $\gamma = 8.96 \times 10^{-7} \textrm{ps}^{-1}$, and low damping, $\gamma = 8.96 \times 10^{-8} \textrm{ps}^{-1}$}. The system is first thermalized at a confinement frequency $\omega_z^{(i)}$ sufficiently far from the critical frequency $\omega_z^{(c)}$ such that the correlation length is small i.e. of the order of the lattice spacing. After that the confining frequency is decreased linearly at a rate, $\omega_z = \omega_z^{(i)}+t(\omega_z^{(f)}-\omega_z^{(i)})/\tau_Q$, such that the system undergoes a transition between a 1-plane and 3-plane structural phases at a rate $1/\tau_Q$. The defects are counted using the method presented in Section \ref{sec:defects}, and in order to obtain the ensemble averaged defect number $\langle n(t) \rangle$ the simulations are carried out $\sim 140$ times for each $\tau_Q$.


Figure \ref{fig:nvst} shows the evolution of the defect density as function of time following phase transition at different quench rates. 
Several dynamical regimes can be seen in the figure. Initially, the system is in the 1-plane phase far from phase transition point, the correlation length is small and the density of defects is large. As $\omega_z$ approaches the critical frequency, the defect density decreases adjusting to the new equilibrium values. This regime crosses over to rapid relaxation, where the density of vortices decreases significantly. According to the KZ theory, this crossover occurs at $t>\hat{t}$. The relaxation continues at a slower pace consistent with the power-law coarsening dynamics. Finally, the relaxation slows further as the system enters the clock phase and the domains are stabilized. \blue{Figure \ref{fig:coarsening} (B) compares the evolution of $n(t)$ with low and high damping at three selected quench rates. For the most part the shape of the $n(t)$ curve does not depend on the damping coefficient, with higher damping resulting in a lag in the dynamics due to the slower response of the system. There is however a significant dependence of $n(t)$ on the damping coefficient at later times, when the system is undergoing coarsening. We observe a higher rate of vortex-antivortex annihilation at lower friction coefficient. Changing friction coefficient can alter the dynamical class of the model, in particular, the limit of zero friction corresponds to energy conserving model.  }


In figure \ref{fig:coarsening} we zoom in into the late time evolution at three different quench rates to verify whether the relaxation in the system follows the coarsening scaling law. The growth of the correlation length due to coarsening in the KT phase \blue{in the high damping regime} is predicted to follow $\xi \sim (t \ln t)^{1/2}$ and consequently the number of defects follow a powerlaw with a logarithmic correction $n \ln n \sim t^{-1}$ \cite{PhysRevE.47.1525}. In figure \ref{fig:coarsening}, one can see that there is a region where the powerlaw with the logarithmic correction is valid.
However, this regime crosses over fairly quickly to a regime of slower relaxation, which we believe is either due to the stabilizing effect of entering into the clock phase or the pinning of the vortices by the discreteness of the lattice.

Finally, figure \ref{fig:lastN} shows the number of defects at the end of the quench at $t=\tau_Q$ as a function of $\tau_Q$. 
\blue{We observe two scaling regimes. At fast quench rates (small $\tau_Q$), more defects are observed in the end of the quench for low friction dynamics. On the other hand, at slower quench rates the final density of defects is higher for high friction simulations. This can be understood intuitively as follows. Lowering the damping coefficient results in effectively higher temperature in the system, lower spatial correlation length and a higher number of topological defects. This explains the higher defect numbers for short quench protocols in the low friction regime. For long quenches, the defects have more time to annihilate and the final low $N_d$ in the low friction regime can be attributed by the higher rate of vortex-antivortex annihilation. For large $\tau_Q$, the fitted powerlaw scaling exponent is -0.9 and -1.07 for the high and low friction respectively. This can be compared to the approximate scaling exponent of -0.72 reported in the numerical study of KZ scaling in the overdamped XY Ginzburg-Landau (GL) field model \cite{Jelic2011}. One should keep in mind, however, that the comparison with previous results is not straightforward, since for KT transition KZ scaling is not a simple powerlaw. In addition, we observe the sensitivity of the scaling to friction and hence the dynamic critical exponent $z$ is likely to be lower than the exponent of the overdamped TDGL model of $z=2$.}

\section{Conclusion}

In this paper we have studied the non-equilibrium dynamics of a planar Coulomb crystals undergoing a structural transition from a 1-planar to 3-planar configurations. The mapping to the Ginzburg-Landau theory reveals that this phase transition corresponds to a transition from a disordered paramagnetic phase to an ordered 6-clock phase with an intermediate KT phase.  We used molecular dynamics simulations to confirm that the KT and the 6-clock phase support stable topological defect structures: vortices, anti-vortices, cross-hairs and asters. 
The density of defects depends on the quench rate of the structural transition as predicted by the Kibble-Zurek theory of defect formation. We have verified that the defect scaling law is consistent with but not identical to the KZ scaling previously observed in numerical simulation of two dimensional XY spin model \cite{Jelic2011}. Moreover, we have observed signatures of coarsening due to defect/anti-defect annihilation, which follows a relaxation powerlaw with a logarithmic correction. 

Our work demonstrates that large planar ion Coulomb crystals can be used as model for studying non-equilibrium statistical physics of vortices in an effectively quasi-two-dimensional system. Such Coulomb crystals with open boundary conditions can be realized in Penning traps. While the focus of this paper was on the system with repulsive Coulomb interactions, similar transition is expected for a lattice with dipolar interactions \cite{PhysRevX.6.031025}. We hope that our work will stimulate research towards experimental study of the predicted topological defects and their rich collective dynamics.

\bibliography{voretexRefs}

\end{document}